\documentclass[10pt,twocolumn]{IEEEtran}
\usepackage[latin9]{inputenc}
\usepackage{amsmath}
\usepackage{amssymb}
\usepackage{graphicx}
\usepackage{setspace}
\usepackage{esint}

\makeatletter
\usepackage{epsfig}\usepackage{amsfonts}\usepackage{latexsym}\usepackage{cite}\usepackage{array}%\usepackage{multicol}
\makeatother

\begin{document}

\title{A Novel Spectrally-Efficient Scheme for Physical Layer Network Coding }

\author{{\small \IEEEauthorblockN{ Ahmed G. Helmy \IEEEauthorrefmark{1},
Tamer Khattab\IEEEauthorrefmark{2}, Mazen O. Hasna\IEEEauthorrefmark{2}}}\\
 {\small {} \IEEEauthorblockA{\IEEEauthorrefmark{1} Department
of Electronics and Electrical Communication, Cairo University, Giza,
Cairo}}\\
{\small {} \IEEEauthorblockA{\IEEEauthorrefmark{2}Electrical
and Computer Engineering, Qatar University, PO Box 2713, Doha, Qatar}}}
\maketitle
\begin{abstract}
In this paper, we propose a novel three-time-slot transmission scheme
combined with an efficient embedded linear channel equalization (ELCE)
technique for the Physical layer Network Coding (PNC). Our transmission
scheme, we achieve about $33\%$ increase in the spectral efficiency
over the conventional two-time-slot scheme while maintaining the same
end-to-end BER performance. We derive an exact expression for the
end-to-end BER of the proposed three-time-slot transmission scheme
combined with the proposed ELCE technique for BPSK transmission. Numerical
results demonstrate that the exact expression for the end-to-end BER
is consistent with the BER simulation results.
\end{abstract}
\setlength{\textfloatsep}{0.4cm} \setlength{\floatsep}{0.4cm} \setlength{\abovecaptionskip}{-0.1cm}

\section{Introduction}

\label{sec_Introduction} Network Coding (PNC) is a relatively new
paradigm in networking which is based on exploiting interference,
instead of avoiding it, to significantly enhance network throughput,
robustness, and security \cite{Yang2010a}. It has been extensively
studied for wired networks and wireless ad-hoc networks \cite{Sagduyu2005,Park2006}.
The concept of physical-layer network coding (PNC) was originally
proposed in \cite{ZhangSept.2006} as a way to exploit network coding
operation \cite{LiFeb.2003,AhlswedeJuly2000} that occurs naturally
in superimposed electromagnetic (EM) waves. The laws of physics show
that when multiple EM waves come together within the same physical
space, they mix. This mixing of EM waves is a form of network coding,
performed by nature. Using PNC in a Two-Way Relay Channel (TWRC) boosts
the system throughput by 100\% \cite{ZhangSept.2006}. 

\begin{figure}
\includegraphics[width=1\columnwidth,height=0.35\columnwidth]{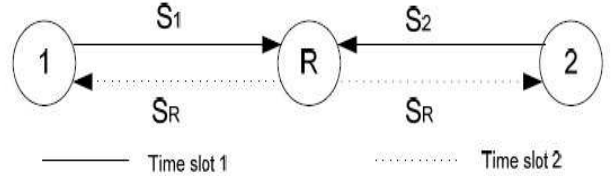}

\caption{Physical Layer Network Coding approach (PNC) \cite{ZhangSept.2006}}

\label{Flo:PNC_net-2} 
\end{figure}

Fig. \ref{Flo:PNC_net-2} illustrates the idea of the concept of network
coding. In the first time slot, nodes 1 and 2 transmit $S_{1}$ and
$S_{2}$ simultaneously to relay R. Relay R deduces $S_{R}$= $S_{1}\oplus S_{2}$
. Then, in the second time slot, relay R broadcasts $S_{R}$ to nodes
1 and 2, where $\oplus$ refers to the XOR operation. 

The main issue in PNC is how relay R deduces $S_{R}$= $S_{1}\oplus S_{2}$
from the superimposed EM waves, which is referred as ``PNC mapping\textquotedblright{}.
Generally, PNC mapping is the process of mapping the received mixed
EM waves plus noise to the desired-network coded signal for forwarding
by the relay to the two end nodes. In general, PNC mapping in not
restricted to the XOR mapping.

In \cite{LuNov.2009}, the authors investigate the Symbol Error Rate
(SER) performance for BPSK and QPSK schemes for two end nodes with
in-phase and orthogonal constellation in AWGN environment. The analysis
assumes perfect channel estimation and takes into consideration the
effect of power control at the two end nodes. The authors use the
Craig's polar coordinate algorithm \cite{CraigNov.1991} to derive
an exact expression for the SER.

Most of the work found in literature assumes that the two received
streams which compose the superimposed EM wave at the relay can be
perfectly resolved and channel-equalized using channel estimates at
the relay based on channel estimation techniques presented in the
literature, such as \cite{JiangJune2010,GaoOct.2009,WangNov.2009}.
Practically, such resolvability assumption contradicts the main principle
of PNC operation which relies on utilizing the natural superposition
of EM waves from both end nodes at the relay to map these signals
into the desired-network coded signal to be forwarded by the relay
to the two end nodes without the separation at the relay.

In this paper, we propose an efficient embedded linear channel equalization
(ELCE) technique to perfectly equalize the channels \textit{without
resolving data streams from each node} using a three-time-slot system
assuming perfect channel estimation at the relay node. In addition
to overcoming the impractical assumption of stream separation, our
 proposed three-time-slot scheme achieves about $33\%$ increase in
spectral efficiency compared to the BPSK transmission presented in
\cite{ZhangSept.2006} while maintaining the same BER performance
of resolvable BPSK and QPSK PNC schemes. The achieved spectral efficiency
lies between the one of BPSK assuming resolvable streams at the relay
node and QPSK assuming resolvable nodes' beams at relay. Finally,
we present an exact analysis for the end-to-end bit-error rate (BER)
expression for the  proposed three-time-slot scheme assuming BPSK
transmission under Rayleigh fading channel.

The rest of this paper is organized as follows: In Section \ref{sec:sysmodel},
we describe our three-time-slot system model. Section \ref{sec:embedded Scheme}
presents our proposed ELCE technique. An exact end-to-end BER expression
for the proposed three-time-slot scheme over the Rayleigh fading channels
is derived in Section \ref{sec:modified-ECE}. In Section \ref{sec:numerical },
we provide the numerical results for the proposed three-time-slot
scheme combined with the ELCE technique and we conclude the paper
in Section \ref{sec:conclusion }.

\section{System Model }

\label{sec:sysmodel}
\begin{figure}
\includegraphics[width=1\columnwidth,height=0.45\columnwidth]{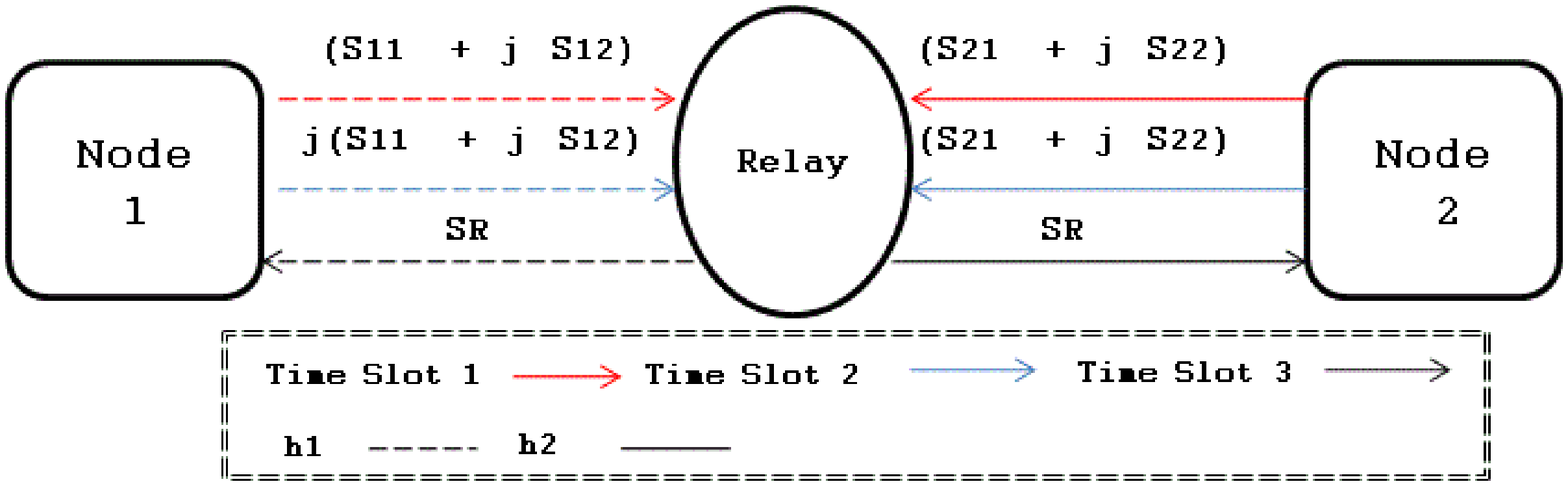}

\caption{Proposed Three-Time-Slot scheme for a communication system using PNC
for TWRC}

\label{Flo:PNC_net} 
\end{figure}

In this section, we introduce our system model
tailored with our proposed three-time-slot scheme for a communication
system using PNC for TWRC. The relay and the users are assumed fully
symbol-synchronized. Channels are assumed to be Rayleigh fading with
channel gains represented as circulary symmetric complex random variables
and the noise is Additive White Gaussian (AWGN) with zero mean. We
also assume that all the channels' state information are available
at the receivers side. As illustrated in Fig. \ref{Flo:PNC_net},
node $1$ and node $2$ send two successive symbols with $90^{o}$
phase difference between them, $i.e.$ node 1 and node 2 will send
$(S_{11}+jS_{12})$ and $(S_{21}+jS_{22})$, respectively, to the
relay node in the first time slot, where $S_{11},\, S_{12}$ and $S_{21},\, S_{22}$
are two successive symbols of node 1 and node 2, respectively. Then,
in the second time slot, node 2 repeats its transmission of $(S_{21}+jS_{22})$,
however, node 1 retransmits a $90^{o}-shifted$ version of its transmission
in the first slot, $i.e.$ it transmits $j(S_{11}+jS_{12})$. The
relay node adopts the ELCE technique, described in Section \ref{sec:embedded Scheme},
followed by a PNC mapping using the superimposed EM waves $Y_{1}$
and $Y_{2}$ received at the relay node in the first two time slots.
In the third slot, the relay transmits the PNC-mapped data $S_{R}$
to the end nodes. Hence, we transmit four symbols in three time slots
which means a $33\%$ spectral efficiency increase over the conventional
two-slot scheme in \cite{ZhangSept.2006}. The two superimposed EM
waves $Y_{1}$ and $Y_{2}$ can be expressed as follows

\begin{align}
Y_{1} & =h_{1}(S_{11}+jS_{12})+h_{2}(S_{21}+jS_{22})+n_{1}\label{y1-1}\\
Y_{2} & =jh_{1}(S_{11}+jS_{12})+h_{2}(S_{21}+jS_{22})+n_{2},\label{y2-1}
\end{align}
where $h_{1}$, $h_{2},$ $n_{1}$, and $n_{2}$ are the channel between
node 1 and the relay, the channel between node 2 and the relay, the
noise at the relay receiver at the first time slot with variance $\sigma_{1}^{2}$,
and the noise at the relay receiver at the second time slot with variance
$\sigma_{2}^{2}$, respectively. We assume that $h_{1}$and $h_{2}$
are block fading channels with constant amplitudes during the full
transmission time ($i.e.$ during the three time slots). We also assume
equal noise variance for $n_{1}$ and $n_{2}$, $i.e.$ $\sigma_{1}^{2}=\sigma_{2}^{2}=\sigma^{2}$. 

\section{Embedded Linear Channel Equalization (ELCE) Technique }

\label{sec:embedded Scheme}In this section, we present the proposed
ELCE technique for perfect channel equalization assuming perfect channel
estimation at the relay and the end nodes. Starting from Eqs. (\ref{y1-1})
and (\ref{y2-1}), we multiply Eqs. (\ref{y2-1}) and (\ref{y1-1})
by $\frac{h_{1}^{*}}{\left\Vert h_{1}\right\Vert ^{2}}$ and $\frac{h_{2}^{*}}{\left\Vert h_{2}\right\Vert ^{2}}$,
respectively to produce $Z_{1}$ and $Z_{2}$, respectively. Hence,
the received-signal vector $\underline{Z}=\left[\begin{array}{c}
Z_{1}\\
Z_{2}
\end{array}\right]$ can be expressed as follows 

\begin{equation}
\underline{Z}=\left[\begin{array}{cc}
j & \frac{h_{1}^{*}h_{2}}{\left\Vert h_{1}\right\Vert ^{2}}\\
\frac{h_{1}h_{2}^{*}}{\left\Vert h_{2}\right\Vert ^{2}} & 1
\end{array}\right]\left[\begin{array}{c}
(S_{11}+jS_{12})\\
(S_{21}+jS_{22})
\end{array}\right]+\left[\begin{array}{c}
\frac{h_{1}^{*}}{\left\Vert h_{1}\right\Vert ^{2}}\, n_{2}\\
\frac{h_{2}^{*}}{\left\Vert h_{2}\right\Vert ^{2}}\, n_{1}
\end{array}\right]
\end{equation}

We construct the channel-equalized-signal vector $\underline{X}=\left[\begin{array}{c}
X_{1}\\
X_{2}
\end{array}\right]$ by left multiplying $\underline{Z}$ by the equalization matrix $\mathbf{\underline{H}}$
which is defined as follows \textbf{
\begin{equation}
\mathbf{\underline{H}}=\mathbf{\frac{1}{2}\mathbf{\left[\vphantom{}\begin{array}{cc}
\left(1+j\right) & 0\\
0 & \left(1-j\right)
\end{array}\right]}\left[\vphantom{}\begin{array}{cc}
\frac{h_{1}h_{2}^{*}}{\left\Vert h_{2}\right\Vert ^{2}} & -j\\
-j & j\,\frac{h_{1}^{*}h_{2}}{\left\Vert h_{1}\right\Vert ^{2}}
\end{array}\right]}\label{w12}
\end{equation}
}therefore, $\underline{X}$ can be expressed as follows

\begin{align}
\underline{X} & =\underline{\mathbf{H}\,}\underline{Z}\nonumber \\
 & =\left[\begin{array}{c}
(S_{21}+jS_{22})\\
(S_{11}+jS_{12})
\end{array}\right]+\frac{1}{2}\left[\begin{array}{c}
\,\,\frac{(1+j)\, h_{2}^{*}}{\left\Vert h_{2}\right\Vert ^{2}}\left(n_{1}-jn_{2}\right)\\
-j\frac{(1-j)\, h_{1}^{*}}{\left\Vert h_{1}\right\Vert ^{2}}\left(n_{1}-n_{2}\right)
\end{array}\right]
\end{align}

The relay node calculates the perfectly channel-equalized combined
signal $X^{ELCE}=X_{1}+X_{2}$ which is equivalent to the superimposed
EM wave at the relay node used to perform the PNC mapping before forwarding
to both end nodes. Hence, the signal $X^{ELCE}$ can be expressed
as follows

\begin{align}
X^{ELCE} & =\left[\mathbf{S_{2}}+\frac{1}{2}\frac{(1+j)\, h_{2}^{*}}{\left\Vert h_{2}\right\Vert ^{2}}\left(n_{1}-jn_{2}\right)\right]\nonumber \\
 & +\left[\mathbf{S_{1}}+\frac{1}{2}\frac{-j(1-j)\, h_{1}^{*}}{\left\Vert h_{1}\right\Vert ^{2}}\left(n_{1}-n_{2}\right)\right]\label{X}
\end{align}
where $\mathbf{S_{1}}=(S_{11}+jS_{12})$ and $\mathbf{S_{2}}=(S_{21}+jS_{22})$

\section{Exact End-to-End BER Performance for the Proposed Three-Time-Slot
Scheme}

\label{sec:modified-ECE}
\begin{figure}
\includegraphics[width=1\columnwidth,height=1\columnwidth]{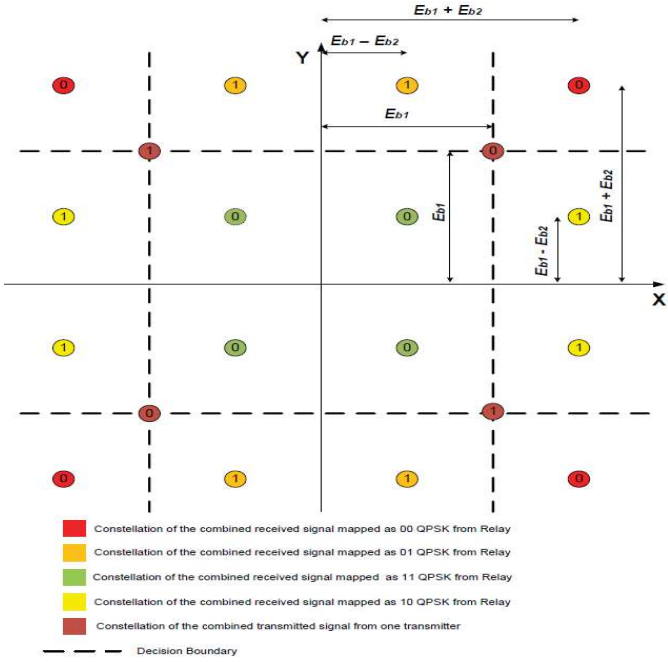}

\caption{The received constellation at the relay node for BPSK signal for the
proposed three-time-slot transmission scheme}

\label{Flo:PNC_net-1-1} 
\end{figure}

In this section, we provide the BER performance
analysis for the  proposed three-time-slot scheme at the relay node
for BPSK modulation scheme at each node. Fig. \ref{Flo:PNC_net-1-1}
shows the received signal constellation at the relay node assuming
that both end nodes use BPSK modulation scheme. We assume that $E_{b_{1}}$
and $E_{b_{2}}$ are the constant bit energy for the BPSK signal generated
from nodes 1 and 2, respectively. Then, each node start performing
the  proposed three-time-slot scheme by combining each two successive
BPSK symbols ($i.e.$ $S_{11},\, S_{12}$ and $S_{21},\, S_{22}$
for node 1 and 2, respectively, with $2$ different possibilities
''$0$ and $1$`` for each symbol) together into one QPSK symbol
($i.e.$ $\mathbf{S_{1}}$ and $\mathbf{S_{2}}$ for node 1 and 2,
respectively, with $4$ different possibilities ''$00$, $01$, $10$,
and $11$'' for each symbol) and transmit it to the relay node. Consequently,
there are sixteen possible symbols in the combined received signal
constellation at the relay node ($i.e.$ for noise-free $X^{ELCE}=\mathbf{S_{1}}+\mathbf{S_{2}}$
). Then, the relay node performs the PNC mapping on the noise-free
$X^{ELCE}$ to construct the QPSK-mapped signal $\mathbf{S_{R}=\mathbf{S_{1}}\oplus\mathbf{S_{2}}}$
and broadcasts it to the end nodes in the third time-slot as shown
in Fig. \ref{Flo:PNC_net}. Since, $\mathbf{S_{R}=\mathbf{S_{1}}\oplus\mathbf{S_{2}=(S_{11}+jS_{12})+(S_{21}+jS_{22})}}$
and $S_{11},\, S_{12}$,$S_{21},$ and $S_{22}$ are BPSK symbols,
hence, each combined symbol $\mathbf{S_{R}}$ at relay node is resulted
by the addition of encoded four bits. However, the relay node maps
$\mathbf{S_{R}}$ to a QPSK PNC-mapped signal to broadcast it to the
end nodes at the third time-slot. We assume that $E_{b_{1}}\geq E_{b_{2}}$,
therefore, we have sixteen decision regions bounded by decision boundaries
$\pm E_{b_{1}}$ for in-phase and quadrature components in the signal
constellation as shown in Fig. \ref{Flo:PNC_net-1-1}. 

To simplify the analysis, we use the Craig's polar coordinate algorithm
\cite{CraigNov.1991} for symbol-error rate (SER) calculation for
AWGN channels. Furthermore, we extend this analysis for the fading
channels by using the instantaneous value of noise variance $\sigma_{N}^{2}\left(|h_{1}|^{2},|h_{2}|^{2}\right)$
which can be proved from Eq. (\ref{X}) to be as $\sigma_{N}^{2}\left(|h_{1}|^{2},|h_{2}|^{2}\right)=\sigma^{2}\left[\frac{1}{|h_{1}|^{2}}+\frac{1}{|h_{2}|^{2}}\right]$
and we can consider $\sigma_{N}^{2}\left(|h_{1}|^{2},|h_{2}|^{2}\right)$
as the new instantaneous noise variance of a zero mean AWGN signal
added to the desired signal $after$ performing the ELCE technique.
Although we apply our analysis to BPSK only, however it can be extended
to higher modulation.

Let $F_{e/0}\left(|h_{1}|^{2},|h_{2}|^{2}\right)$ denotes the instantaneous
probability of symbol error in the PNC mapping process at the relay
due to the noise effect assuming that the noise-free PNC-mapped signal
is ``\textbf{0}'' ($i.e.$ $S_{R}=0$), and $F_{e/1}\left(|h_{1}|^{2},|h_{2}|^{2}\right)$
denotes the instantaneous probability of symbol error in the PNC mapping
process at the relay due to the noise effect assuming that the noise-free
PNC-mapped signal is ``\textbf{1}'' ($i.e.$ $S_{R}=1$), where
$|h_{1}|^{2}$ and $|h_{2}|^{2}$ are the channel gains for $h_{1}$and
$h_{2}$, respectively.

Using Craig's polar coordinate algorithm \cite{CraigNov.1991}, we
develop the instantaneous expressions for $F_{e/0}\left(|h_{1}|^{2},|h_{2}|^{2}\right)$
and $F_{e/1}\left(|h_{1}|^{2},|h_{2}|^{2}\right)$ exploiting the
previous definition of $\sigma_{N}^{2}\left(h_{1}|^{2},|h_{2}|^{2}\right)$.
To simplify the notation, we denote $\sigma_{N}^{2}=\sigma_{N}^{2}\left(h_{1}|^{2},|h_{2}|^{2}\right)=\sigma^{2}\left[\frac{\gamma_{1}+\gamma_{2}}{\gamma_{1}\gamma_{2}}\right]$,
$\gamma_{1}=|h_{1}|^{2}$, $\gamma_{2}=|h_{2}|^{2},$ $F_{e/0}^{inst}\left(\gamma_{1},\gamma_{2}\right)=F_{e/0}\left(|h_{1}|^{2},|h_{2}|^{2}\right)$,
and $F_{e/1}^{inst}\left(\gamma_{1},\gamma_{2}\right)=F_{e/1}\left(|h_{1}|^{2},|h_{2}|^{2}\right)$.
Consequently, the average probability of symbol error over the fading
channel given that symbol ``\textbf{0}'' was transmitted $P_{e/0}^{Symbol}$
and the average probability of symbol error over the fading channel
given that symbol ``\textbf{1}'' was transmitted $P_{e/1}^{Symbol}$
can be expressed as follows

\begin{align}
P_{e/0}^{Symbol} & =\int_{\gamma_{1}}\int_{\gamma_{2}}F_{e/0}^{inst}\left(\gamma_{1},\gamma_{2}\right)\, f_{\gamma_{1}}(\gamma_{1})f_{\gamma_{2}}(\gamma_{2})d\gamma_{1}d\gamma_{2}\label{pe0}\\
P_{e/1}^{Symbol} & =\int_{\gamma_{1}}\int_{\gamma_{2}}F_{e/1}^{inst}\left(\gamma_{1},\gamma_{2}\right)\, f_{\gamma_{1}}(\gamma_{1})f_{\gamma_{2}}(\gamma_{2})d\gamma_{1}d\gamma_{2}\label{pe1}
\end{align}
where $f_{\gamma_{1}}(\gamma_{1})$ and $f_{\gamma_{2}}(\gamma_{2})$
are the probability density function (PDF) of the channel gains of
$\gamma_{1}$ and $\gamma_{2}$, respectively. From \cite{CraigNov.1991},
we derive $F_{e/0}^{inst}\left(\gamma_{1},\gamma_{2}\right)$ and
$F_{e/1}^{inst}\left(\gamma_{1},\gamma_{2}\right)$ as follows

\begin{align}
F_{e/0}^{inst}\left(\gamma_{1},\gamma_{2}\right) & =\frac{1}{\pi}\sum_{k=1}^{K}\int_{0}^{\phi_{k}}exp\left\{ -\frac{A_{k}^{2}}{2\sigma_{N}^{2}\, sin^{2}\theta}\right\} d\theta\nonumber \\
 & =\frac{1}{\pi}\sum_{k=1}^{K}\int_{0}^{\phi_{k}}exp\left\{ -\frac{A_{k}^{2}\,\gamma_{1}\gamma_{2}}{2\sigma^{2}(\gamma_{1}+\gamma_{2})\, sin^{2}\theta}\right\} d\theta\label{f0}\\
F_{e/1}^{inst}\left(\gamma_{1},\gamma_{2}\right) & =\frac{1}{\pi}\sum_{l=1}^{L}\int_{0}^{\phi_{l}}exp\left\{ -\frac{A_{l}^{2}}{2\sigma_{N}^{2}\, sin^{2}\theta}\right\} d\theta\nonumber \\
 & =\frac{1}{\pi}\sum_{l=1}^{L}\int_{0}^{\phi_{l}}exp\left\{ -\frac{A_{l}^{2}\,\gamma_{1}\gamma_{2}}{2\sigma^{2}(\gamma_{1}+\gamma_{2})\, sin^{2}\theta}\right\} d\theta\label{f1}
\end{align}
where $K$ and $L$ are the number of all possible error regions assuming
that the noise-free PNC-mapped symbol ``\textbf{0}'' was transmitted
and the number of all possible error regions assuming that the noise-free
PNC-mapped symbol ``\textbf{1}'' was transmitted, respectively.
In addition, $\phi_{k}$ and $\phi_{l}$ are the scanning angle for
each of the error regions of the noise-free PNC-mapped symbol ``\textbf{0}''
and the scanning angle for each of the error regions of the noise-free
PNC-mapped symbol ``\textbf{1}'', respectively. The parameters $A_{k}^{2}/2\sigma^{2}$,
and $A_{l}^{2}/2\sigma^{2}$ are the received symbol energy projected
on the decision boundary divided by the noise density for each of
the error regions of the noise-free PNC-mapped symbol ``\textbf{0}''
and the received symbol energy projected on the decision boundary
divided by the noise density for each of the error regions of the
noise-free PNC-mapped symbol ``\textbf{1}'', respectively. All of
these parameters depend on the signal constellation received at the
relay node which will be shown later on for our probability of symbol
error derivation in Sections \ref{sub:symbol_0} and \ref{sub:symbol_1}.

Let $\Gamma$ denotes a new random variable which is defined as $\Gamma=\frac{\gamma_{1}\gamma_{2}}{\gamma_{1}+\gamma_{2}}$,
we apply a random variable transformation to deduce the PDF of $\Gamma$;
namely $f_{\Gamma}(\Gamma)$ in terms of the PDFs of $\gamma_{1}$
and $\gamma_{2}$. Using Eqs. (\ref{f0}) and (\ref{f1}) and employing
the definition of $\Gamma$, Eqs. (\ref{pe0}) and (\ref{pe1}) can
be expressed as follows

\begin{spacing}{1.3}
\begin{align*}
P_{e/0}^{Symbol} & =\int_{\Gamma}F_{e/0}^{inst}\left(\Gamma\right)\, f_{\Gamma}(\Gamma)d\Gamma\\
 & =\frac{1}{\pi}\sum_{k=1}^{K}\int_{0}^{\phi_{k}}\left[\int_{0}^{\infty}exp\left\{ -\frac{A_{k}^{2}\,\Gamma}{2\sigma^{2}sin^{2}\theta}\right\} f_{\Gamma}(\Gamma)d\Gamma\right]d\theta\\
P_{e/1}^{Symbol} & =\int_{\Gamma}F_{e/1}^{inst}\left(\Gamma\right)\, f_{\Gamma}(\Gamma)d\Gamma\\
 & =\frac{1}{\pi}\sum_{l=1}^{L}\int_{0}^{\phi_{l}}\left[\int_{0}^{\infty}exp\left\{ -\frac{A_{l}^{2}\,\Gamma}{2\sigma^{2}sin^{2}\theta}\right\} f_{\Gamma}(\Gamma)d\Gamma\right]d\theta
\end{align*}
where $F_{e/0}^{inst}\left(\Gamma\right)$ and $F_{e/1}^{inst}\left(\Gamma\right)$
are the instantaneous probability of symbol error as a function of
$\Gamma$ for ``\textbf{0}'' and ``\textbf{1}'' noise-free PNC-mapped
symbols, respectively. The inner integral (in square brackets) is
in the form of a Laplace transform with respect to the variable $\Gamma$.
Since the moment generating function (MGF) of $\Gamma$ {[}i.e., $M_{\Gamma}(s)=\int_{0}^{\infty}e^{s\Gamma}f_{\Gamma}(\Gamma)\, d\Gamma$
{]} is the Laplace transform of $f_{\Gamma}(\Gamma)$ with the exponent
reversed in sign. Consequently, $P_{e/0}^{Symbol}$ and $P_{e/1}^{Symbol}$
expressions can be rewritten as follows\cite{Simon2005} 
\begin{eqnarray}
P_{e/0}^{Symbol} & = & \frac{1}{\pi}\sum_{k=1}^{K}\int_{0}^{\phi_{k}}M_{\Gamma}\left\{ -\frac{A_{k}^{2}}{2\sigma^{2}sin^{2}\theta}\right\} \, d\theta\label{pe_0}\\
P_{e/1}^{Symbol} & = & \frac{1}{\pi}\sum_{l=1}^{L}\int_{0}^{\phi_{l}}M_{\Gamma}\left\{ -\frac{A_{l}^{2}}{2\sigma^{2}sin^{2}\theta}\right\} \, d\theta\label{pe_1}
\end{eqnarray}

\end{spacing}

Eqs. (\ref{pe_0}) and (\ref{pe_1}) are considered the general forms
used to evaluate the average probability of symbol error for any binary
signal constellation over an arbitrary distribution of fading channels
$h_{1}$ and $h_{2}$ and consequently $\gamma_{1}$ and $\gamma_{2}$.
For the Rayleigh fading channel, $\gamma_{1}$ and $\gamma_{2}$ are
exponentially distributed with average $\overline{\gamma}_{1}$ and
$\overline{\gamma}_{2}$, respectively. For the sake of simplicity,
we assume that $\overline{\gamma}_{1}=\overline{\gamma}_{2}=\overline{\gamma}$.
Using the definition of the MGF of $\Gamma$ $M_{\Gamma}\left(s\right)=_{2}F_{1}\left(1,\,2;\,\frac{3}{2};\,-\frac{\overline{\gamma}}{4}s\right)$
expressed in (\cite{Hasna2003}, Eq. 20), the general forms in Eqs.
(\ref{pe_0}) and (\ref{pe_1}) can be rewritten for the Rayleigh
fading channels after some mathematical manipulations as follows

\begin{align}
P_{e/0}^{Symbol} & =\frac{1}{\pi}\sum_{k=1}^{K}\int_{0}^{\phi_{k}}{}_{2}F_{1}\left(1,\,2;\,\frac{3}{2};\,\frac{\overline{\gamma}}{4}\frac{A_{k}^{2}}{2\sigma^{2}sin^{2}\theta}\right)\, d\theta\label{pe_1-0}\\
P_{e/1}^{Symbol} & =\frac{1}{\pi}\sum_{l=1}^{L}\int_{0}^{\phi_{l}}{}_{2}F_{1}\left(1,\,2;\,\frac{3}{2};\,\frac{\overline{\gamma}}{4}\frac{A_{l}^{2}}{2\sigma^{2}sin^{2}\theta}\right)\, d\theta\label{pe_1-1}
\end{align}
where $_{2}F_{1}(a,b;c;z)$ is the hypergeometric function for the
parameters $a$, $b$, $c$, and $z$. The integral in Eqs. (\ref{pe_1-0})
and (\ref{pe_1-1}) can be evaluated numerically using any approximation
technique such as Gauss Quadrature Numerical Integration Method. Let
$P_{relay}^{S}$ denotes the total average probability of symbol error
at the relay node over an arbitrary fading channel distributions assuming
equally probable binary signal transmission. $P_{relay}^{S}$ can
be expressed as follows

\begin{equation}
P_{relay}^{S}=\frac{1}{2}\left(P_{e/0}^{Symbol}+P_{e/1}^{Symbol}\right)\label{pe-2}
\end{equation}

Without loss of generality and assuming Gray coded bit mapping at
both end nodes. Since, $\mathbf{S_{R}=\mathbf{S_{1}}\oplus\mathbf{S_{2}=(S_{11}+jS_{12})+(S_{21}+jS_{22})}}$
and $S_{11},\, S_{12}$,$S_{21},$ and $S_{22}$ are BPSK symbols,
hence, each combined symbol $\mathbf{S_{R}}$ at relay node is resulted
by the addition of Gray encoded four bits that differ by only one
bit from the adjacent combined symbol, $i.e.$ if the noise causes
the constellation to cross the decision boundary, only one out of
the four bits, combined to generate the symbol received at relay node,
will be in error. Consequently, the relation between the BER $P_{relay}^{b}$
and the SER for the combined symbol at the relay node will be approximately
as follows

\begin{equation}
P_{relay}^{b}\thickapprox P_{relay}^{S}/4\label{bit-er}
\end{equation}

Then, the end-to-end BER from node 1 to node 2, $P_{1\rightarrow2}$,
is defined as the BER between the data transmitted from node 1 and
decoded at node 2 as follows

\begin{align}
P_{1\rightarrow2} & =1-(1-P_{relay}^{b})(1-P_{r,2})\nonumber \\
 & =P_{relay}^{b}+P_{r,2}-P_{r,2}P_{relay}^{b}\label{Pt2}
\end{align}
with $P_{r,2}=Q\left(\frac{E_{R}}{\sigma}\sqrt{\gamma}\right)$ indicates
the BER caused by the data transmission from the relay to node 2,
where $E_{R}$ and $Q(x)$ are the constant bit energy used by the
relay node to transmit the QPSK PNC-mapped signal to the end nodes,
and $Q(x)=\frac{1}{\sqrt{2\pi}}\int_{x}^{\infty}e^{-\lambda^{2}/2}d\lambda$.
By the new definition of the $Q-function$ presented in \cite{Simon2005},
the BER $P_{r,2}$ value for the Rayleigh fading channel, for a value
of AWGN variance $\sigma^{2}$ and channel gain $\overline{\gamma}$,
will be as follows

\[
P_{r,2}=\frac{1}{2}\left(1-\sqrt{\frac{E_{R}^{2}\overline{\gamma}/2\sigma^{2}}{1+E_{R}^{2}\overline{\gamma}/2\sigma^{2}}}\right)
\]

Similarly, the end-to-end BER from node 2 to node 1, $P_{2\rightarrow1}$,
is defined as the BER between the data transmitted from node 2 and
decoded at node 1 as follows

\begin{align}
P_{2\rightarrow1} & =1-(1-P_{relay}^{b})(1-P_{r,1})\nonumber \\
 & =P_{relay}^{b}+P_{r,1}-P_{r,1}P_{relay}^{b}\label{Pt1}
\end{align}
with $P_{r,1}=Q\left(\frac{E_{R}}{\sigma}\sqrt{\gamma}\right)$ indicates
the BER caused by the data transmission from the relay to node 1.
Also, the BER $P_{r,1}$ value for the Rayleigh fading channel, for
a value of AWGN variance $\sigma^{2}$ and channel gain $\overline{\gamma}$,
will be as follows

\[
P_{r,1}=\frac{1}{2}\left(1-\sqrt{\frac{E_{R}^{2}\overline{\gamma}/2\sigma^{2}}{1+E_{R}^{2}\overline{\gamma}/2\sigma^{2}}}\right)
\]

Finally, the overall end-to-end BER for an equal given channel gain
$\overline{\gamma}$ and AWGN variance $\sigma^{2}$ is obtained,
using Eqs. (\ref{Pt1}) and (\ref{Pt2}) and the definitions of $P_{r,1}$
and $P_{r,2}$, as follows

\begin{equation}
P_{overall}^{e}=\frac{1}{2}\left(P_{1\rightarrow2}+P_{2\rightarrow1}\right)\label{Ptot}
\end{equation}

In the next subsections, we derive the SER in the PNC mapping process
at the relay due to the noise effect assuming that the noise-free
PNC-mapped combined symbols are ``\textbf{0}'' and ``\textbf{1}''.
Recalling Eqs. (\ref{pe_1-0}) and (\ref{pe_1-1}) on the signal constellation
shown in Fig. \ref{Flo:PNC_net-1-1-1-1-2} and Fig. \ref{Flo:PNC_net-1-1-1-1},
respectively, we calculate the total BER at relay node $P_{relay}^{b}$
using Eq. (\ref{bit-er}). We use signal constellation to derive the
values of of the controlling parameters $\phi_{k},$$\phi_{l},$$A_{k}^{2}$,
and $A_{l}^{2}$ for both cases of the noise-free PNC-mapped combined
symbols ``\textbf{0}'' and ``\textbf{1}''. Once we compute $P_{relay}^{b}$,
the overall end-to-end BER $P_{overall}^{e}$ can be evaluated by
Eq (\ref{Ptot}) for given channel parameters $\overline{\gamma}$
and $\sigma^{2}$.

\subsection{SER of the PNC-mapped Combined Symbol ``0'' \label{sub:symbol_0}}

\begin{figure}
\begin{centering}
\includegraphics[width=1\columnwidth,height=0.6\columnwidth]{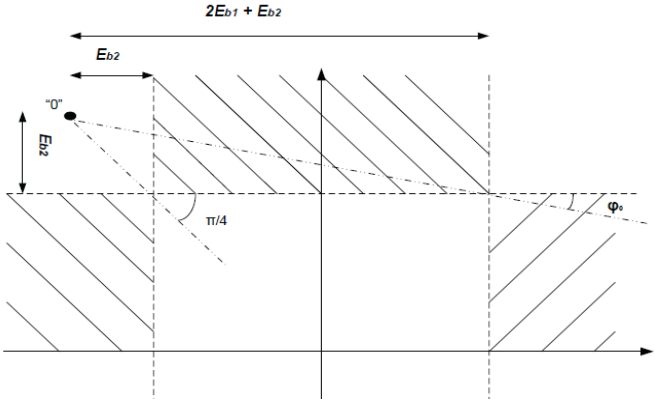} 
\par\end{centering}

\caption{Decision boundaries and decoding for the PNC-mapped combined symbol\textbf{
0}}

\label{Flo:PNC_net-1-1-1-1-2} 
\end{figure}

To understand the decoding process for the PNC-mapped combined symbol
\textbf{``0}'', we use the signal constellation geometry in Fig.
\ref{Flo:PNC_net-1-1}. As shown in Fig. \ref{Flo:PNC_net-1-1-1-1-2},
the channel-equalized symbol $X^{ELCE}$ is considered an error in
this case when it is located in the shaded regions. The expression
of $P_{e/0}^{Symbol}$ can be derived, using Eq. (\ref{pe_1-0}),
as follows

\begin{align}
P_{e/0}^{Symbol} & =\frac{1}{\pi}\int_{0}^{\pi/2}{}_{2}F_{1}\left(1,\,2;\,\frac{3}{2};\,\frac{\overline{\gamma}}{4}\frac{\left[E_{b_{2}}\right]^{2}}{2\sigma^{2}sin^{2}\theta}\right)\, d\theta\nonumber \\
 & -\frac{1}{\pi}\int_{0}^{\pi/2}{}_{2}F_{1}\left(1,\,2;\,\frac{3}{2};\,\frac{\overline{\gamma}}{4}\frac{\left[2E_{b_{1}}+E_{b_{2}}\right]^{2}}{2\sigma^{2}sin^{2}\theta}\right)\, d\theta\nonumber \\
 & +\frac{1}{\pi}\int_{0}^{\pi/2}{}_{2}F_{1}\left(1,\,2;\,\frac{3}{2};\,\frac{\overline{\gamma}}{4}\frac{\left[E_{b_{2}}\right]^{2}}{2\sigma^{2}sin^{2}\theta}\right)\, d\theta\nonumber \\
 & -\frac{1}{\pi}\int_{0}^{\pi/4}{}_{2}F_{1}\left(1,\,2;\,\frac{3}{2};\,\frac{\overline{\gamma}}{4}\frac{\left[E_{b_{2}}\right]^{2}}{2\sigma^{2}sin^{2}\theta}\right)\, d\theta\nonumber \\
 & -\frac{1}{\pi}\int_{0}^{\pi/4}{}_{2}F_{1}\left(1,\,2;\,\frac{3}{2};\,\frac{\overline{\gamma}}{4}\frac{\left[E_{b_{2}}\right]^{2}}{2\sigma^{2}sin^{2}\theta}\right)\, d\theta\nonumber \\
 & +\frac{1}{\pi}\int_{0}^{\pi/2-\varphi_{0}}{}_{2}F_{1}\left(1,\,2;\,\frac{3}{2};\,\frac{\overline{\gamma}}{4}\frac{\left[2E_{b_{1}}+E_{b_{2}}\right]^{2}}{2\sigma^{2}sin^{2}\theta}\right)\, d\theta\nonumber \\
 & +\frac{1}{\pi}\int_{0}^{\varphi_{0}}{}_{2}F_{1}\left(1,\,2;\,\frac{3}{2};\,\frac{\overline{\gamma}}{4}\frac{\left[E_{b_{2}}\right]^{2}}{2\sigma^{2}sin^{2}\theta}\right)\, d\theta\label{SER_0}
\end{align}
where $\varphi_{0}=tan^{-1}\left\{ \frac{E_{b_{2}}}{2E_{b_{1}}+E_{b_{2}}}\right\} $

\subsection{SER of the PNC-mapped Combined Symbol ``1''\label{sub:symbol_1}}

\begin{figure}
\begin{centering}
\includegraphics[width=1\columnwidth,height=0.6\columnwidth]{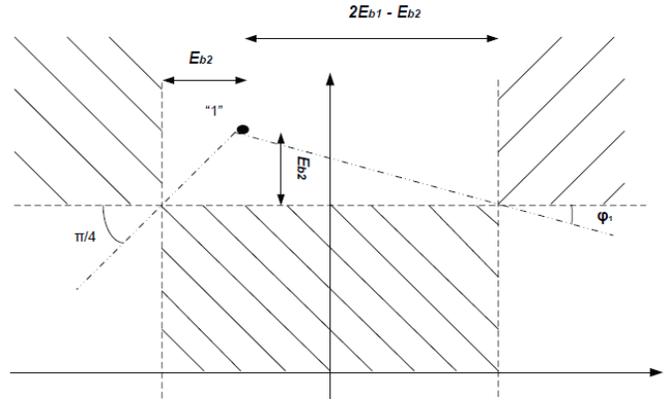} 
\par\end{centering}

\caption{Decision boundaries and decoding for the PNC-mapped combined symbol
\textbf{1}}

\label{Flo:PNC_net-1-1-1-1} 
\end{figure}

To understand the decoding process for the symbol ``\textbf{1}'',
we use the signal constellation geometry in Fig. \ref{Flo:PNC_net-1-1}.
As shown in Fig. \ref{Flo:PNC_net-1-1-1-1}, the channel-equalized
symbol $X^{ELCE}$ is considered an error in this case when it is
located in the shaded regions. The expression of $P_{e/1}^{Symbol}$
can be derived, using Eq. (\ref{pe_1-1}), as follows

\begin{align}
P_{e/1}^{Symbol} & =\frac{1}{\pi}\int_{0}^{\pi/2}{}_{2}F_{1}\left(1,\,2;\,\frac{3}{2};\,\frac{\overline{\gamma}}{4}\frac{\left[E_{b_{2}}\right]^{2}}{2\sigma^{2}sin^{2}\theta}\right)\, d\theta\nonumber \\
 & +\frac{1}{\pi}\int_{0}^{\pi/2}{}_{2}F_{1}\left(1,\,2;\,\frac{3}{2};\,\frac{\overline{\gamma}}{4}\frac{\left[2E_{b_{1}}-E_{b_{2}}\right]^{2}}{2\sigma^{2}sin^{2}\theta}\right)\, d\theta\nonumber \\
 & +\frac{1}{\pi}\int_{0}^{\pi/2}{}_{2}F_{1}\left(1,\,2;\,\frac{3}{2};\,\frac{\overline{\gamma}}{4}\frac{\left[E_{b_{2}}\right]^{2}}{2\sigma^{2}sin^{2}\theta}\right)\, d\theta\nonumber \\
 & -\frac{1}{\pi}\int_{0}^{\pi/4}{}_{2}F_{1}\left(1,\,2;\,\frac{3}{2};\,\frac{\overline{\gamma}}{4}\frac{\left[E_{b_{2}}\right]^{2}}{2\sigma^{2}sin^{2}\theta}\right)\, d\theta\nonumber \\
 & -\frac{1}{\pi}\int_{0}^{\pi/4}{}_{2}F_{1}\left(1,\,2;\,\frac{3}{2};\,\frac{\overline{\gamma}}{4}\frac{\left[E_{b_{2}}\right]^{2}}{2\sigma^{2}sin^{2}\theta}\right)\, d\theta\nonumber \\
 & -\frac{1}{\pi}\int_{0}^{\pi/2-\varphi_{1}}{}_{2}F_{1}\left(1,\,2;\,\frac{3}{2};\,\frac{\overline{\gamma}}{4}\frac{\left[2E_{b_{1}}-E_{b_{2}}\right]^{2}}{2\sigma^{2}sin^{2}\theta}\right)\, d\theta\nonumber \\
 & -\frac{1}{\pi}\int_{0}^{\varphi_{1}}{}_{2}F_{1}\left(1,\,2;\,\frac{3}{2};\,\frac{\overline{\gamma}}{4}\frac{\left[E_{b_{2}}\right]^{2}}{2\sigma^{2}sin^{2}\theta}\right)\, d\theta\label{SER_1}
\end{align}
where $\varphi_{1}=tan^{-1}\left\{ \frac{E_{b_{2}}}{2E_{b_{1}}-E_{b_{2}}}\right\} $

\section{Numerical Results}

\begin{onehalfspace}
\label{sec:numerical }In this section, we present our numerical results
for our  proposed three-time-slot scheme in Fig. \ref{Flo:PNC_net}
and the end-to-end BER performance analysis for the received constellation
shown in Fig. \ref{Flo:PNC_net-1-1}. Assume zero-mean white Gaussian
noise and consider slow Rayleigh fading channels with flat amplitudes,
we consider $E_{b_{1}}=4$, $E_{b_{2}}=2$, $E_{R}=1$, and $\overline{\gamma}=20dB$.
\end{onehalfspace}

\begin{figure}
\includegraphics[width=1\columnwidth,height=0.85\columnwidth]{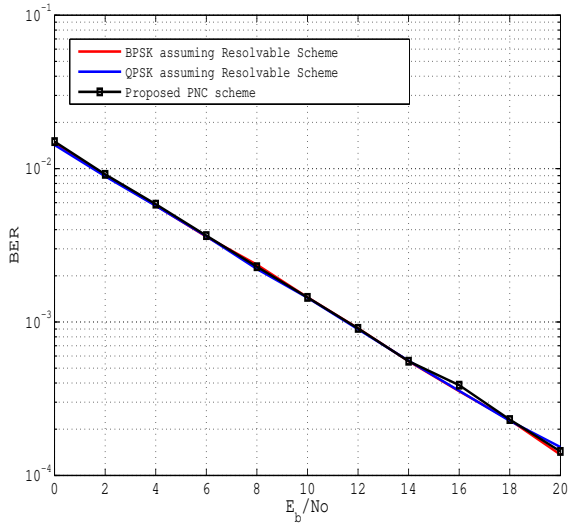}

\caption{BER performance comparison between the proposed three-time-slot transmission
scheme and resolvable BPSK and QPSK}

\label{Flo:PNC_net-1-1-1-1-1} 
\end{figure}

Fig. \ref{Flo:PNC_net-1-1-1-1-1} depicts the end-to-end BER performance
comparison between the  proposed three-time-slot scheme and the resolvable
BPSK and QPSK. Fig. \ref{Flo:PNC_net-1-1-1-1-1} demonstrates that
the proposed scheme achieves the same end-to-end BER performance of
the resolvable BPSK and QPSK with higher spectral efficiency.

\begin{figure}
\includegraphics[width=1\columnwidth,height=0.85\columnwidth]{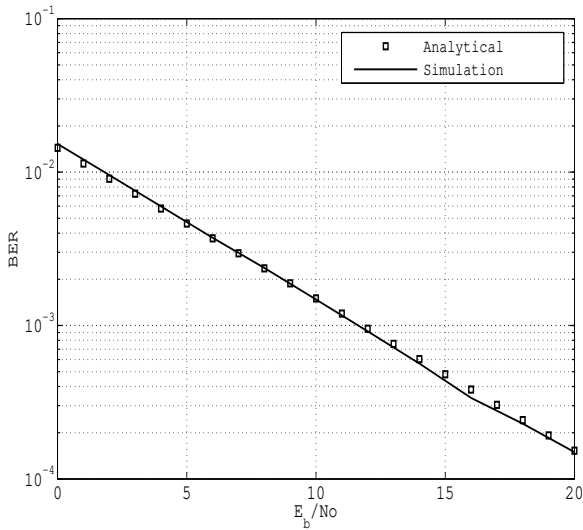}

\caption{Comparison between analytical and simulation results for the proposed
three-time-slot transmission scheme}

\label{Flo:PNC_net-1-1-1-1-1-1} 
\end{figure}

In Fig. \ref{Flo:PNC_net-1-1-1-1-1-1}, we compare between the the
simulation results of end-to-end BER for the  proposed three-time-slot
scheme and the others from the analytical expression for BER numerically
calculated from Eq. (\ref{Ptot}). Fig. \ref{Flo:PNC_net-1-1-1-1-1-1}
demonstrates that the analytical expression for the end-to-end BER
is consistent with the simulation results.

\section{Conclusion}

\label{sec:conclusion } In this paper, we proposed a novel three-time-slot
transmission scheme combined with an efficient ELCE technique. Using
such three-time-slot transmission scheme, we achieved about $33\%$
increase in the spectral efficiency over the conventional two-time-slot
scheme with the same end-to-end BER performance as shown in our numerical
results. In addition, we provided an exact expression for the end-to-end
BER for the proposed three-time-slot scheme in case of BPSK transmission.
Numerical results demonstrate that the provided exact analytical expression
of the end-to-end BER of the proposed three-time-slot scheme is almost
consistent with the BER simulation results.

\bibliographystyle{IEEEtran}
\bibliography{IEEEabrv,PNC_Survey,qnrf_ref,Journal1,Sparse_IA,reference}

\end{document}